# Comparison of Capital Costs per Route-Kilometre in Urban Rail

Bent Flyvbjerg*, Nils Bruzelius** and Bert van Wee***
* Department of Development and Planning
Aalborg University
Fibigerstraede 13
9220 Aalborg
Denmark
tel: +45 9940 7216
fax: +45 9815 3788
e-mail: flyvbjerg@plan.aau.dk
and
Faculty of Technology, Policy and Management
Delft University of Technology
Jaffalaan 5
2628 BX Delft
The Netherlands

** Independent consultant and university lecturer on transportation and planning
tel: +46 46 151354
fax: +46 708 151356
e-mail: nils.bruzelius@euromail.se

*** Faculty of Technology, Policy and Management
Delft University of Technology
Jaffalaan 5
2628 BX Delft
tel: +31 15 2787186
fax: +31 15 2782719
e-mail: bertw@tudelft.nl

*Because of the prominent position of urban rail in reducing urban transport-related problems, such as congestion and air pollution, insights into the costs of possible new urban rail projects is very relevant for those involved with cost estimations, policy makers, cost-benefit analysts, and other target groups. Knowledge of the differences in costs per kilometre,*

*including explanations of differences and their breakdowns is currently lacking in the literature. This paper aims to provide a first stage insight into how cost per kilometre varies across urban rail projects. The methodology applied is a simple cost comparison across projects where the data collected are comparable. We conclude that capital costs per route-kilometre of urban rail vary highly between projects. Looking at European projects and excluding outliers, the total capital costs per route-kilometre (including stations and rolling stock) lie mainly between US$50-100 million (2002 prices). Including US projects, the range is US$50-150 million. The main reasons for the high variation in the route-kilometre costs are differences between projects as regards the ratio of underground to above-ground construction, ground conditions, station spacing, type of rolling stock, environmental and safety constraints and labour costs. We warn, however, that the observations used to reach the conclusions are too few to obtain results with statistical significance. Our results must therefore be seen as a first step towards collecting more data so that a more succinct statistical analysis can be conducted. Another conclusion is therefore that this area has future research potential.*

## 1. Introduction

Most large urban areas in western countries, as well as in developing countries, face major transport problems. Congestion levels are high, demand for parking capacity increasingly exceeds supply, concentrations of pollutants exceed WHO levels, and about 20 percent of the European Union's population suffer from unhealthy noise levels (European Commission, 1996). Whilst the numbers of fatalities are generally decreasing, they are still at a high and often politically unacceptable level. In Europe, as in other regions, solutions for these problems involve both push measures – such as the congestion charge in London and tolls in some Norwegian cities – and pull measures, such as improving public transport, as well as, sometimes, a combination of push and pull measures, such as in London where revenue from the congestion charge is used to improve public transport. Public transport infrastructure improvements are high on the agenda of pull measures in many cities, partly inspired by success stories such as Karlsruhe (Germany), or Strasbourg (France). Because of intensive land use in large urban areas in Europe, especially in central urban areas, improved public transport often involves underground systems.

An important question in the discussion on possible new urban rail projects is: will the costs be compensated for by the benefits? Currently, cost-benefit analysis (CBA) is the most recognised *ex ante* evaluation method for transport infrastructure projects (Hayashi and Morisugi, 2000). However, scientific discussions and guidelines (e.g. those in the UK and in the Netherlands) focus more on the valuation of benefits than the quality of cost estimates. This is surprising, as estimates of costs are highly uncertain, with cost over-runs being the rule rather than the exception. Rail projects on average perform even worse than roads, with an average cost over-run of 45 percent (Flyvbjerg et al., 2003).

The costs of infrastructure projects have recently attracted considerable attention in the literature (e.g. Flyvbjerg et al., 2003; Odeck, 2004; see Van Wee, 2007 for a review of the literature) but this work mainly focuses on cost *over-runs* and explanatory factors, and very little on unit costs, such as costs per kilometre. We did find some references to costs per kilometre for rail projects, but they only examined cost changes over time (e.g. Cox and Love, 1995), and not how unit costs vary across projects.

This paper aims to provide further insights into the unit costs of urban rail projects. The quality of urban rail projects as well as the CBA of those projects could be improved from such insights, and the decision making might also benefit. More specifically, we believe the results may prove useful to those interested in the costs of urban rail, including planners, cost engineers, project appraisers and investors. This paper therefore attempts to answer the following questions:

1. What are the average costs per kilometre of urban metro rail projects constructed in the recent past?
2. Which factors contribute to the explanation of differences observed in unit costs?
3. Into which categories can the urban rail project costs be broken down?

We consider our research as a first step towards a more succinct statistical analysis of these factors, given that there is little data at our disposal and no statistical inference can be conducted. Another aim of the paper is to reflect on the possible implications of our findings for further research in this area.

Section 2 describes briefly the method used. Section 3 describes the data; Section 4 the results and Section 5 contains a summary of the main conclusions.

## 2. Methodology

The approach we used in this study is a simple comparison of the unit costs of urban rail projects that have been built in Europe, supplemented by some data from the US and elsewhere. In general, studies of this type should encompass a succinct statistical assessment where inferences can be made with respect to the significance of the results derived. However, because the number of projects considered in the present study was low due to unavailability of data, we resorted to a simpler approach, which gives insights into the cost differences that may be useful for those involved with cost estimations, planners and decision makers. As we later recommend, this area deserves future research encompassing more data gathering and statistical analysis to reach firm conclusions. Hence, our study must be seen as a first stage in such an endeavour.

## 3. Data

Data for the present analysis was used in several stages. In the first stage cost data for urban rail systems and lines were collected as part of a wider research into the costs and cost over-runs of transport infrastructure projects. In the second stage a decision was made as to which types of projects would be further considered. We decided to focus on urban rail projects in densely populated areas in Europe as it is seldom an area of focus as far as implementation costs are concerned. Further, focusing on this area gives a homogenous type of project that enables a worthwhile comparison of costs. Projects were selected that were wholly or at least partly underground as tunnelling is a key element in most European rail infrastructure projects. Secondly, European projects were considered more relevant than projects from the USA or other nations as the focus of the research is on Europe, although examples of non-European nations are included. Thirdly, data on light rail and commuter rail have not been included.

The quality of data gathered is also an issue to consider. The data collection exercise showed that capital cost data, as found in the records of the owners and managers of urban rail and in the reports produced by consultants and researchers, are highly uneven regarding specifications and quality. It is often not clear which items are included in a specific cost measure, for instance whether or not costs include land, rolling stock, management costs, taxes, etc. Moreover, it is sometimes unclear whether costs are given in constant or current prices and to which year a given cost figure pertains. As a result one has to be careful not to make the apples-and-oranges error when comparing cost figures for one system with those for another, including benchmarking one system with data from others.

Taken together, such problems typically make the data on costs that are immediately available for a given urban rail system or line unsuited for comparative studies. Therefore, before data on unit capital costs were used for comparison and benchmarking purpose, three steps were taken:

1. Costs are compared for similar systems i.e. urban rail, European projects and at least partly underground projects, to ensure homogeneity in the data
2. Costs were expressed in constant (real) prices, using construction cost indices to discount costs to the same level (year).
3. Costs calculated in different currencies were converted into the same currency, typically US$ or €, by applying the appropriate exchange rates.

In (2) and (3) above, we first took the estimate of the cost in a given currency for a specific year. That cost was then updated to 2002 using the construction cost index for that country using OECD (see OECD, 1997). Finally the 2002 average exchange rate was used to translate the figures into US$. Where a construction cost index was not available, the index of the US was used. The reason for using the US index is that most of the projects are without a local index in Latin American countries. US construction costs is assumed to be more relevant than European costs.

Our cost estimates reflect all the costs accrued until the moment the project opens, i.e. the first year the project was completed and brought into use. Costs that were incurred after the opening, e.g. due to minor shortcomings, were excluded. This is because it is difficult to distinguish between repairs of faults that occurred in the construction phases from maintenances. Furthermore, data availability did not make that distinction possible. It should be noted that the word 'costs' is used from the perspective of the client and may thus include the contractor's margin. One could argue that a better word would be 'prices', because we focus on the prices paid for the projects, not the real costs – costs would imply ignoring profits and losses. However, to be in line with the literature we use the word 'costs' instead of 'prices'. To be clear, costs as used here is what was incurred or paid by the building authorities to realize the project.

There are however, additional factors other than the ones described above which may help to explain the differences between projects. These include differences related to time periods and the country/region of construction. Construction costs can easily differ due to changes in the real costs of the construction industry and local market circumstances during the construction period. In addition one could argue that the level of cost over-runs could be included: a project could be more expensive than necessary simply because of cost over-runs. But this is not as straightforward as it seems: initial cost estimates are often biased for strategic reasons, and therefore unusable as 'realistic' costs for which the projects could have been constructed (Flyvbjerg, 2007b). We are the first to admit that our approach to data collection must be considered as a first step towards acquiring sufficient data for this type of study. We recommend therefore that these two issues, the impact of time periods and the country/region of construction be subjects for further research.

## 4. Results

Given the methodology discussed above, which basically compares the implementation costs across a set of homogenous rail projects in Europe, and given the data described in section 3, we can now address the three questions posed in section (1).

### 4.1 Do Capital Costs per Route-kilometre differ between Metros?

The question posed in this section is what the average costs per kilometre of urban rail projects constructed in the recent past are. We distinguish between the unit costs for the total project and the unit costs for stations only.

Table 1 presents the results for three groups of European projects. Projects were firstly selected applying the criteria as presented in section 2. Secondly, as we aimed to give a good impression of the range in the values of variables, we selected projects selectively rather than randomly. Total capital costs per route-kilometre include stations. It should also be mentioned that the costs quoted for the Copenhagen metro cover 21 kilometres, of which only 11 had been completed at the time of the survey; turn-out per-kilometre costs for the 21 kilometres may therefore differ from the figure in the table. The indicated costs for the Toulouse VAL Line A extension are also an estimate.

Table 1 illustrates well a general characteristic of metro systems: capital costs per kilometre vary significantly from project to project, here the lowest unit costs are for the Hannover U-Bahn and the highest for the London Jubilee Line extension. The Hannover U-Bahn has an even lower unit cost than the Madrid extension - well-known for its low costs - which can be explained by the fact that only 17 percent of the metro in Hannover is underground.

**Table 1. Capital costs per route-kilometre for selected urban rail projects.**

|  | Opening year | Length km | Vertical segregation | Number of stops. Stop spacing km | Capital costs (million) | Costs/km (million) | Cost/km (million) 2002-US$ |
|---|---|---|---|---|---|---|---|
| Copenhagen Metro Phases 1-3 | 2002-07 | 21 | 48% tunnel 52% elevated | 22 1.0 | DKK 11,400 | DKK 542.9 | 69.8 |
| London Jubilee Line extension | 1999 | 16 | 78% tunnel 22% at ground level | NA NA | GBP 3,600 | GBP 225 | 329.9 |
| Madrid Extension 1995-99 | 1999 | 56.3 | 68% tunnel 32% at ground level | 38 1.5 | NA | US$22.8 | 26.7 |
| Toulouse VAL Line A | 1993 | 9.7 | 90% tunnel 10% elevated | 15 0.6 | FRF 3,700 | FRF 381.4 | 60.9 |
| Toulouse VAL Line A extension | 2004 | 2.2 | NA | 3 0.7 | € 187.5 | € 85.2 | 81.1 |
| Marseille Lines 1-2 | 1977-92 | 19.6 | 80% tunnel 12% elevated 8% at ground level | 24 0.8 | FRF 6,343 | FRF 323.7 | 59.1 |
| Lille VAL RT | 1988 | 29 | 75% tunnel 25% above | NA 0.7 | FRF 8,900 | FRF 306.9 | 56.0 |
| Lyon Ligne D | 1991-97 | 14 | NA | 15 0.9 | FRF 7,300 | FRF 521.4 | 79.5 |
| Paris Meteor Phase 1 | 1998 | 7.2 | NA | 7 1.0 | US$ 1,419 | US$ 197.1 | 220.0 |
| Marseille Line 1 extension | 2006 | 2.5 | NA | 4 0.6 | € 175,4 | € 70.2 | 68.8 |
| Toulouse VAL Line B | 2007 | 15 | NA | 20 0.8 | € 968 | € 64.5 | 63.2 |
| London Victoria Line | 1968-69 | 15.8 | 100% tunnel | NA 1.3 | € 740.5 | € 46.9 | 63.1 |
| Vienna Stage 1 | 1984 | NA | NA | NA NA | NA | € 70 | 94.2* |
| Berlin U-Bahn | NA | 4.6 | 100% tunnel | 5 0.9 | US$ 275 | US$ 59.8 | 88.3* |
| Hannover U-Bahn | NA | 69.0 | 17% tunnel | 110 0.6 | US$ 750 | US$ 10.9 | 16.1* |
| Hannover U-Bahn extension | NA | 2.8 | 100% tunnel | NA NA | US$ 108 | US$ 38.5 | 56.9* |
| Turin Metro Phase 1 | 2005 | 9.6 | 100% tunnel | 15 0.6 | GBP 442 | GBP 40 | 71.7 |

NA: Not available.
*Exchange rate for the indicated year was not available to convert to local currency. Construction cost index has been applied directly in EUR/US.
Note: Taxes were included for Madrid (16%). Status of taxes for London, Vienna, Berlin, Hannover, and Turin was unknown. Other projects: taxes not included.

We now broaden the scope to projects outside the EU. Tables 2 and 3 present data for 12 metros in the USA and other non-European countries.

**Table 2. Capital costs per route-kilometre for six metros in the USA.**

|  | Washington, DC Metro | Atlanta MARTA | Baltimore Metro Section A&B | Los Angeles North Hollywood extension | Atlanta North Line extension | San Francisco BART Airport extension |
|---|---|---|---|---|---|---|
| Opening year | 1985 | 1986 | 1983 | 2000 | 2000-03 | 2002 |
| Length km | 97.3 | 43.1 | 12.2 | 10.1 | 3.7 | 14.0 |
| % in tunnel, elevated, at ground level | 57% tunnel | 42% tunnel | 56% tunnel | NA | NA | NA |
| Number of stops | 57 | 26 | 9 | 3 | 2 | 4 |
| Stop spacing km | 1.7 | 1.7 | 1.4 | 3.4 | 1.85 | 3.5 |
| Capital costs m | 7,968 | 2,720 | 1,289 | 1,311 | 463.2 | 1,510.2 |
| Costs/km m US$ | 81.9 | 63.1 | 105.7 | 129.8 | 125.2 | 107.9 |
| *Costs/km m 2002-US$* | *114.3* | *88.0* | *147.5* | *131.6* | *126.9* | *109.4* |

NA: Not available.
Note: Status of taxes unknown.

**Table 3. Capital costs per route-kilometre for six metros in Asian and Latin American nations**

|  | Singapore | Seoul | Calcutta | Mexico City Line B | Caracas Line 3 | Santiago Line 5 extension |
|---|---|---|---|---|---|---|
| Opening year | NA | NA | NA | 2000 | 1994 | 2000 |
| Length km | 67 | 116.5 | 16.5 | 23.7 | 4.4 | 2.8 |
| % in tunnel, elevated, at ground level | 30% tunnel | 80% tunnel | 95% tunnel | 25% tunnel, 20% elevated 55% at ground level | 100% tunnel | 100% tunnel |
| Number of stops | NA | NA | NA | 21 | 4 | 3 |
| Stop spacing km | 1.6 | 1.1 | 1.0 | 1.1 | 1.1 | 0.9 |
| Capital costs m | 2,500 | 5,240 | 684 | 970 | 372 | 197 |
| Costs/km m US$ | 37.3 | 45.0 | 41 | 40.9 | 84.5 | 70.4 |
| *Costs/km m 2002-US$** | *54.5* | *65.8* | *59.9* | *43.8* | *98.4* | *71.8* |

NA: Not available.
* US construction cost index has been applied.
Note: Taxes included for Mexico City (15%), Caracas (16%) and Santiago (6%). Status of taxes unknown for Singapore, Seoul and Calcutta.

In addition to the unit cost figures in tables 1 to 3, the International Association for Public Transport, UITP, provided us with the following figures (see table 4).

**Table 4. UITP data for selected projects**

| Athens | US$156 million per route-kilometre |
|---|---|
| Cairo | US$109 " |
| Frankfurt a.m. | US$108 " |
| Lisbon | US$118 " |

Unfortunately, it was not possible to obtain information regarding the year in which these costs were calculated or to which specific metro lines they refer. A figure from UITP for the London Jubilee Line of US$375 million per kilometre compared to our figure of US$330 million per kilometre (see table 1) suggests that the UITP figures may be slightly on the high side.

Nevertheless, these figures and the figures in tables 2 and 3 confirm the initial impression from table 1 of large variations between projects in per-kilometre capital costs. The Hannover U-Bahn, Madrid metro extension and the London Jubilee line are notable outliers, even when seen in the context of the larger number of projects from outside Europe presented in tables 2 and 3. The American metros in table 2 tend to have higher per-kilometre costs than their counterparts both in Europe and in Asian and Latin American countries.

Excluding the outliers, the US metros and the UITP figures, the total capital costs per kilometre for metros is in general between US$50-100 million. Including the US and UITP figures the range is US$50-150 million per kilometre.

As mentioned earlier, the available data include the costs of stations in the total capital costs. The data typically do not allow the separating out of stations for independent cost analysis. However, Pickrell (1985), in one of the most thorough studies of unit construction costs that exists, estimated a cost for underground stations of US$40 million, elevated stations of US$23 million and at-grade stations of US$10 million (1983 dollars, see table 5). It should be noted, however, that the study was carried out in 1983 and considers North American metros only. As with the other estimates of metro unit costs, these figures show large variability.

**Table 5. Estimates of unit construction costs relating to stations.**

| Component | Unit construction costs Millions of 1983 dollars |
|---|---|
| Underground stations | 40 |
| Elevated stations | 23 |
| Ground level stations | 10 |
| Two-track km, in tunnel: | |
| Including stations | 85 |
| Excluding stations | 64 |
| Two-track km, elevated: | |
| Including stations | 34 |
| Excluding stations | 24 |
| Two-track km, at ground level: | |
| Including stations | 19 |
| Excluding stations | 14 |

Source: Pickrell 1985, p. 58.

Table 5 shows that including station costs increases project costs by 33-42%. In other words, station costs have a share of 25-29 % of the overall project costs. In addition, table 5 shows that elevated stations are more than twice as expensive as ground level stations, and that underground stations are about four times as expensive as ground level stations. We present table 5 only to give data on the share of stations in the total costs rather than to give insights into station costs expressed in 2002 prices, and therefore have used Pickrell's (1985) original data.

### 4.2 Explanations of Cost Variations

In this section we aim to answer the second research question: which factors contribute the most to explaining differences in unit costs?

Our data did not contain enough information to be able to derive specific answers. Therefore, in attempting to answer this question we relied on previous literature.

A key factor in explaining the cost variations between projects is the cost and complexity of establishing the right of way (establishing the corridor, including buying the land if needed). Costs may be as low as US$10 million per kilometre if an at-grade right of way is available for conversion for free, but can rise to over US$200 million per kilometre for an underground railway in a difficult urban terrain with troublesome geology, and high costs for land acquisition and clearance for stations, relocation and compensation for existing businesses and residences, etc. (Halcrow Fox 2000).

A further factor is whether stations are below, on or above the ground. Halcrow Fox (2000) recently found that underground construction for new metros is 4 to 6 times as costly and elevated construction 2 to 2.5 times as costly as at-grade construction (see table 6).

**Table 6. Typical costs for new-build metros.**

| Vertical Alignment | All-in costs, US$ million per route-km (2000 prices) | Ratio |
| --- | --- | --- |
| Ground level | 15-30 | 1 |
| Elevated | 30-75 | 2-2.5 |
| Underground | 60-180 | 4-6 |

Source: Halcrow Fox (2000).

In a study for the World Bank other factors that strongly impact on route-kilometre costs were found to be the quality of management, whether a new system is constructed or lines are progressively added to an existing system, and the extent of utilities diversions, environmental constraints and safety requirements (BB&J Consult, 2000). Unfortunately no information is available with respect to the importance of private versus public finance. In addition, costs may be influenced by specific market conditions. For example, if the demand for infrastructure construction is higher than the available (regional or national) capacity, profit margins could be relatively high, leading to higher 'costs' as defined above. In addition, financial markets may have an impact on the prices. Both subjects, private versus public finance as well as market conditions are interesting areas for further research.

In a study carried out by BB&J Consult (2000) an analysis was made of the substantial cost difference between the recent extension of the Madrid metro and extensions of the metro systems in Mexico City, Santiago and Caracas. The analysis relates mainly to tunnel sections. To summarise, as far as infrastructure is concerned the findings of BB&J are that the low Madrid costs can be explained by:

1. General reasons, which account for savings of 15-20 million US$/km. These general reasons include strong political commitment, a highly experienced project management team, and contract procurement not based on the cheapest bid.
2. Specific reasons related to civil works, accounting for savings of up to 10 million US$/km. These include the use of a twin track single tunnel and the Earth Pressure Boring Method (EPBM), strong geotechnical supervision, monitoring and standardised station designs.

3. Specific cost reduction in equipment, accounting for up to 10 million US$/km savings. These are explained by no air conditioning at stations, limited uninterrupted power supply systems, overhead rigid rail or catenary for trains, ATP (automatic train protection) and ATO (automatic train operation), tested technology for signalling and telecommunications, and conventional steel wheels.
4. Specific cost reduction in design, supervision and management accounting for savings of 1 to 5 million US$/km, including short construction time, small project management team, limited technical assistance and the possibility of exploiting scale economics.

### 4.3 Breakdown of Capital Costs

In this section we answer the third research question which is: what are the categories that urban rail project costs can be broken down into?

For six of the metros mentioned in section 4 the available data allow a breakdown of capital costs by subsystem. In addition, such a breakdown was available for four metros not mentioned earlier (San Francisco BART, Chicago CTA, Boston MBTA and Santiago Line 5). For these four projects the total unit costs per kilometre were not available and they were therefore not included in section 4.

Table 7 shows the breakdown of capital costs for five US metros. Table 8 exhibits a similar breakdown for the recent Madrid metro extension and the extensions of the three metros in Caracas, Mexico City and Santiago, employing a somewhat different itemisation of costs than that used for the US metros. For Madrid, the Arganda extension of line 9 has been excluded, because it is mainly a surface line and because the focus of the costs analysis quoted was on tunnelling. For the US metros, it was possible to separate out the capital costs for stations, while this was not the case for the non-US metros. Note that, contrary to previous results, the costs of vehicles/rolling stock are included, because the original source included them. For the purpose of presenting the results, giving insights into the breakdown of costs, this inclusion is not problematic.

Civil works for guide-ways and stations is the largest cost item for all the metros, except for the Mexico City Line B. For the latter, the relatively low percentage of costs for civil works and the high percentage of costs for rolling stock are explained by the fact that only 25 percent of the line is underground (see table 3).

For the US metros, the costs of engineering, management and tests plus the costs of rolling stock are the second most costly items. For the non-US metros, rolling stock is in second place, followed by equipment.

Other cost items, like track, power, traffic control etc., each account for less than ten percent of the total costs for the metros studied.

Finally, it should be noted that the percentage of costs within each item varies substantially across projects. Such variation can be explained by differences between projects in the percentage of underground construction, the capacity of rolling stock, the extent to which goods and services - for instance land and management - were available for free, etc.

**Table 7. Breakdown of metro capital costs by subsystem for five US metros.**

| Subsystem | San Francisco BART (%) | Atlanta MARTA Phase A (%) | Baltimore MTA Phase I (%) | Chicago CTA O'Hare (%) | Boston MBTA Red Line South (%) |
|---|---|---|---|---|---|
| Land | 7 | 9 | 2 | 0 | 11 |
| Guideway | 37 | 33 | 25 | 20 | 15 |
| Stations | 19 | 20 | 30 | 28 | 33 |
| Trackwork | 3 | 2 | 2 | 7 | 7 |
| Power | 2 | 1 | 2 | 5 | 6 |
| Control | 4 | 2 | 4 | 8 | 7 |
| Facilities | 2 | 3 | 2 | 4 | 0 |
| Eng./Mgt./Test | 14 | 23 | 24 | 8 | 6 |
| Vehicles | 12 | 7 | 9 | 20 | 15 |
| *Total* | 100 | 100 | 100 | 100 | 100 |

Source: Federal Transit Authority 1992.

**Table 8. Breakdown of metro capital costs by item for five metro extensions in Madrid, Caracas, Mexico City and Santiago.**

| Item | Mexico City Line B (%) | Caracas Line 3 (%) | Santiago Line 5 (%) | Santiago Line 5 extension (%) | Madrid extension excl. Arganda (%) |
|---|---|---|---|---|---|
| Civil works for tunnels only | 24.5 | 32.8 | 36.4 | 40.5 | 54.6 |
| Equipment | 18.0 | 24.2 | 16.6 | 13.9 | 14.2 |
| Rolling stock | 36.2 | 15.7 | 24.8 | 21.4 | 15.4 |
| Design and supervision | 3.4 | 3.6 | 5.5 | 10.3 | 1.9 |
| Track | 5.3 | 2.8 | 6.3 | 4.3 | 3.5 |
| Power | 5.2 | 8.9 | 4.4 | 2.8 | 2.4 |
| Signalling and communications | 4.1 | 6.8 | 4.3 | 5.2 | 2.7 |
| Station equipment | 0.2 | 2.3 | 0.7 | 0.4 | 1.7 |
| Escalators and lifts | 0.3 | 2.1 | 0.2 | 0.9 | 2.7 |
| Passenger toll equipment | 0.5 | 0.7 | 0.5 | 0.2 | 0.3 |
| Workshop equipment | 2.3 | - | 0.3 | - | 0.6 |
| *Total* | 100 | 100 | 100 | 100 | 100 |

Source: BB&J Consult (2000)

## 5. Conclusions, implications, and further research

In this paper, we have shown that capital costs per route-kilometre in urban rail vary a great deal both in Europe and elsewhere in the world. Our approach has been a simple cost comparison across projects with no statistical testing. The variations observed are large, involve different types worldwide and imply that researchers, planners and decision makers need to worry about their causes.
From our study, we can derive several conclusions which must be seen as preliminary given the data assessed. The first conclusion is that capital costs per kilometre of metro line vary substantially between cities, between metro systems and between metro lines within the same city and system. Looking at European projects and excluding outliers, the total capital costs per route-kilometre (including stations and rolling stock) lie mainly in the interval US$50-100 million (2002 prices), stations having a share of about 25-30%.

The implications of this conclusion are linked to a conclusion from previous studies on related subjects. Previous research has concluded that available data on the capital costs for transport infrastructure projects show that cost estimates are often highly inaccurate and that inaccuracies are particularly high for urban rail (Flyvbjerg, 2007a; Flyvbjerg, Holm and Buhl 2003, 2004; Flyvbjerg, Bruzelius, and Rothergatter, 2003). Our conclusion in this respect, combined with the conclusions of earlier research, point to the possibility of using the results of *ex post* analyses as presented in this paper for *ex ante* forecasting of the costs of urban rail projects. Given the variability in the estimated unit costs, we do not see them as being of use for estimating the expected costs for proposed new projects. The expected costs of such projects will have to be estimated in a conventional way, i.e. by using the project design giving information on quantities such as lengths, width, characteristics of bridges etc. and unit prices. However, 'our' unit prices could be used as a check on the reasonableness of such conventional cost estimates, i.e. they could be used as the basis for posing questions about cost estimates in case there are large differences in the costs estimated by the two approaches.

The second conclusion is that the main reasons for the high variation in the route-kilometre costs are firstly due to differences between project characteristics, as regards the ratio of underground to above-ground construction and ground conditions. Secondly, they may be due to management characteristics. Thirdly, civil works for guide-ways carries the largest costs for most metros, followed by station costs and the costs of engineering, management and tests.

We consider this research as a first step; it certainly does not provide the final answers to the three research questions. Further research is needed to serve project managers, CBA researchers and other target groups. This further research could focus on the following subjects:

- The extension of the database, allowing for statistical analysis and modelling.
- The extension of the scope of the database, resulting in the inclusion of other (urban or non-urban) rail projects, in particular light rail and conventional rail,
- A more sophisticated analysis of cost-explaining factors that are not project specific, such as changes in the real costs of the construction industry related to technological change, change in construction methods, and changes in prices for inputs depending on market factors.
- Establishing benchmarks for unit costs: establishing these costs in such a way that they are suitable for comparison is a methodologically under-developed area. There is still a lot of work to be done before final benchmarks can be said to have been achieved in a statistically valid fashion.
- Converting costs to one currency and one year. It must be noted that, excluding the Copenhagen metro, the conversion and discounting of costs into 2002-US dollars is a first estimate which should be done in more detail if one or more of the projects was to be used as a more detailed benchmark for a new project.
- In depth research, including interviews with project managers and others involved in the projects. Such research might focus on best practices, but also on 'failures', although the latter category might be more difficult because of the dependence on collaboration of the persons involved.

- Research into the link between costs per kilometre of infrastructure and cost overruns: whether there is a significant relationship between both, and if so: why?
- The development of practically applicable guidelines for CBA research, related to the costs of urban rail infrastructure.
- Research into private versus public finance and into the impacts of specific markets (mainly: for construction capacity, for physical inputs, and for finance).

Related to the first two recommendations, an extended database of higher quality will result. This database could very well be used for the method of 'reference class forecasting' as initially developed by Lovallo and Kahneman (2003) in theory, and worked out in practice for transport infrastructure projects by Flyvbjerg (2006). Reference class forecasting consists of taking a so-called "outside view" on the particular project being forecast. The outside view is established on the basis of information from a class of similar projects. The outside view does not try to forecast the specific uncertain events that will affect the particular project, but instead places the project in a statistical distribution of outcomes from this class of reference projects.

## Acknowledgement

The authors wish to thank three anonymous reviewers for their valuable comments.